\documentclass[]{aa}
\renewcommand{\linenumbers}{}
\renewcommand{\nolinenumbers}{}
\renewcommand{\runninglinenumbers}{}

\usepackage{graphicx}
\usepackage{txfonts}
\usepackage{subcaption}         
\usepackage{lscape}             
\usepackage{placeins}           
\usepackage{orcidlink}

\usepackage{natbib,twoopt}

\begin{document} 
\nolinenumbers

   \title{The Competing Influence of Ram Pressure and Tidal Interaction in NGC 2276}

   \author{L. Matijević
          \inst{1,2}\thanks{First author: lmatijevic@pmf.hr}\orcidlink{0009-0004-2049-7701}
          \and
          N. Tomičić\inst{1}\thanks{Corresponding author: ntomicic.phy@pmf.hr}\orcidlink{0000-0002-8238-9210}
          \and
          A. Marasco\inst{2}\orcidlink{0000-0002-5655-6054}
          \and
          A. Ignesti\inst{2}\orcidlink{0000-0003-1581-0092}
          \and
          A. E. Lassen\inst{2}\orcidlink{0000-0003-3575-8316}
          \and
          R. Smith\inst{3,4}\orcidlink{0000-0001-5303-6830}
          \and
          P. Sell\inst{5}\orcidlink{0000-0003-1771-5531}
          \and
          I. D. Roberts\inst{6,7}\orcidlink{0000-0002-0692-0911}
          \and
          A. Zezas\inst{8,9}\orcidlink{0000-0001-8952-676X}
          \and
          K. Anastasopoulou\inst{10,11}\orcidlink{0000-0002-3240-6609}
          \and
          P. Kotoulas\inst{12}
          \and
          R. Bašić\inst{1}
          }

   \institute{Department of Physics, Faculty of Science, University of Zagreb, Bijenička Cesta 32, 10000 Zagreb, Croatia
         \and
         INAF–Padova Astronomical Observatory, Vicolo dell’Osservatorio 5, 35122 Padova, Italy
         \and
         Departamento de Fisica, Universidad Tecnica Federico Santa Maria, Avenida España 1680, Valparaíso, Chile
         \and
         Millenium Nucleus for Galaxies (MINGAL)
         \and
         Georgia Institute of Technology, North Avenue, Atlanta, GA 30332
         \and
         Waterloo Centre for Astrophysics \& Department of Physics \& Astronomy, University of Waterloo, 200 University Ave W, Waterloo, ON N2L 3G1, Canada
         \and
         Leiden Observatory, Leiden University, PO Box 9513, 2300 RA Leiden, The Netherlands
         \and
         Physics Department \& Institute of Theoretical and Computational Physics, University of Crete, GR 71003 Heraklion, Greece
         \and
         Institute of Astrophysics, Foundation for Research and Technology-Hellas, GR 71110 Heraklion, Greece
         \and
         Center for Astrophysics $|$ Harvard \& Smithsonian, 60 Garden Street, Cambridge, MA 02138, USA
         \and
         Istituto Nazionale di Astrofisica (INAF) – Osservatorio Astronomico di Palermo, Piazza del Parlamento 1, 90134 Palermo, Italy
         \and
         Department of Physics, University of Florida, Gainesville, FL 32611-8440, USA
             }

 
  \abstract
   {The evolution of galaxies in groups is profoundly influenced by a variety of physical processes, with ram pressure and tidal interactions playing pivotal roles in shaping their structural and evolutionary pathways. The relative influence of these two processes is still debated in groups compared to clusters, as ram pressure is less understood there. We study NGC 2276, a nearby galaxy (z $\approx$ 0.0079) where the dominant process is still an open question.}
   {We examine the distribution of stellar populations in NGC 2276 using multiwavelength data to assess potential evidence of tidal interactions and ram-pressure stripping.}
   {We present the first HST WFC3/UVIS images of NGC 2276, and use them to investigate the distribution of stellar populations across the disk of NGC 2276, where we assume that the bluer broadband filters mainly trace younger stellar populations, while the redder filters trace mainly older stellar populations. Furthermore, by comparing HST images with maps of H$\alpha$ emission from Calar Alto's PMAS/PPAK integral field unit (IFU) and near-IR maps from Spitzer's IRAC, we identify arm-like overdensity features that trace the spiral structure of this galaxy and tracked the variation of their pitch angle with radius.}
   {Our results indicate that the distribution of the stellar populations is asymmetrical. The youngest stellar populations (up to $\sim$100 Myr) show higher concentration on the leading side of the galaxy and are more diffuse on the trailing side, consistent with gas compression due to ram-pressure. This asymmetry is visible in the red filters as well. We also show that the average pitch angles of the overdensity features increase with galactocentric distance. Our findings are consistent with the fact that ram pressure is the leading mechanism for the peculiar morphology of NGC 2276, but do not exclude the possibility that tidal interactions could have played a role.}
   {}

   \keywords{galaxies: interactions --
             galaxies: evolution --
             galaxies: kinematics and dynamics --
             galaxies: ISM --
             galaxies: groups: individual: NGC 2300 --
             galaxies: individual: NGC 2276 
               }

   \titlerunning{NGC 2276 morphology and environmental effects}

   \maketitle
%

\section{Introduction}

    Galaxy clusters and groups provide excellent conditions for studying how the host environment affects galaxy evolution. Due to the high galaxy number density, the large galaxy-to-galaxy relative speed, and the presence of a hot environmental plasma filling the volume of the intra- group/cluster medium (IGM), group and cluster galaxies can be affected by both gravitational (e.g. tidal stripping,  harassment) and hydrodynamical (e.g. ram-pressure stripping, RPS) environmental effects \citep[e.g.][]{kennicutt, poggianti, schaefer, boselli}.

\begin{table*}[h!]
\caption{General properties of NGC 2276}
\label{tab:params}
\centering
\begin{tabular}{lllll}
\hline
\hline
Parameter                   & Value                                         & Reference                                     \\ \hline
R.A.                        & 7$^\mathrm{h}$27$^\mathrm{m}$14.2$^\mathrm{s}$  & HST WFC3/UVIS filters                                 \\
Dec.                        & +85°45$'$16.2$''$                               & HST WFC3/UVIS filters                                 \\
Inclination                 & 20$\pm$10 {[}°{]}                             & \citet{tomicic18}                                     \\
P.A.                        & 244 {[}°{]}                                   & \citet{tomicic18}                                     \\
systemic velocity           & 2416 {[}km/s{]}                               & \citet{tomicic18}                                     \\
D$_L$                       & 28.5 {[}Mpc{]}                                & Luminosity distance to NGC 2300 group \citep{fadda}   \\
redshift                    & 0.0079                                        & H$\alpha$ emission line \citep{tomicic18}             \\
log$_{10}$(M$_*$/M$_\odot$) & 10.59$\pm$0.2                                 & IRAC 1 and IRAC 2 using \citet{querejeta}             \\
                            & 10.73$\pm$0.1                                 & IRAC 1 using \citet{querejeta}                        \\
IGM mean density            & $\sim5\times10^{-27} \mathrm{g\,cm^{-3}}$     & \citet{mulchaey, rasmussen}                   \\ \hline                                                                                 
\end{tabular}
\end{table*}
    
    The effects of ram pressure have been extensively studied in clusters \citep[e.g.][]{dressler, poggianti,boselli}, and galaxy groups \citep[e.g.][]{roberts21, vulcani, kolcu}, as well as through dedicated numerical simulations \citep[e.g.][]{vollmer, kapferer, tonnesen, akerman23}. Similarly, the gravitational influence through the tidal interaction was also studied both through numerical experiments \citep[e.g.][]{toomre}, and observationally \citep[e.g.][]{lambas, alonso}. Spatially resolved, multiwavelength studies are key to map the main galaxy properties - such as stellar and gas mass, star formation rate (SFR) and metallicity - that in turn can provide key information on the role of the environment in regulating galaxy evolution \citep[e.g.][]{goods03, cosmos07}.  
    
    While specific mechanisms are expected to be caused by specific environmental regimes \citep[e.g.][]{marasco, vulcani}, it is not unusual for a galaxy to be influenced by multiple environmental effects at the same time. The study of these highly disturbed systems can be challenging, but it can also shed light on the complex physics of the interplay between galaxies and their environments.  
    
\begin{figure}[h!]
        \centering
        \includegraphics[width = \hsize]{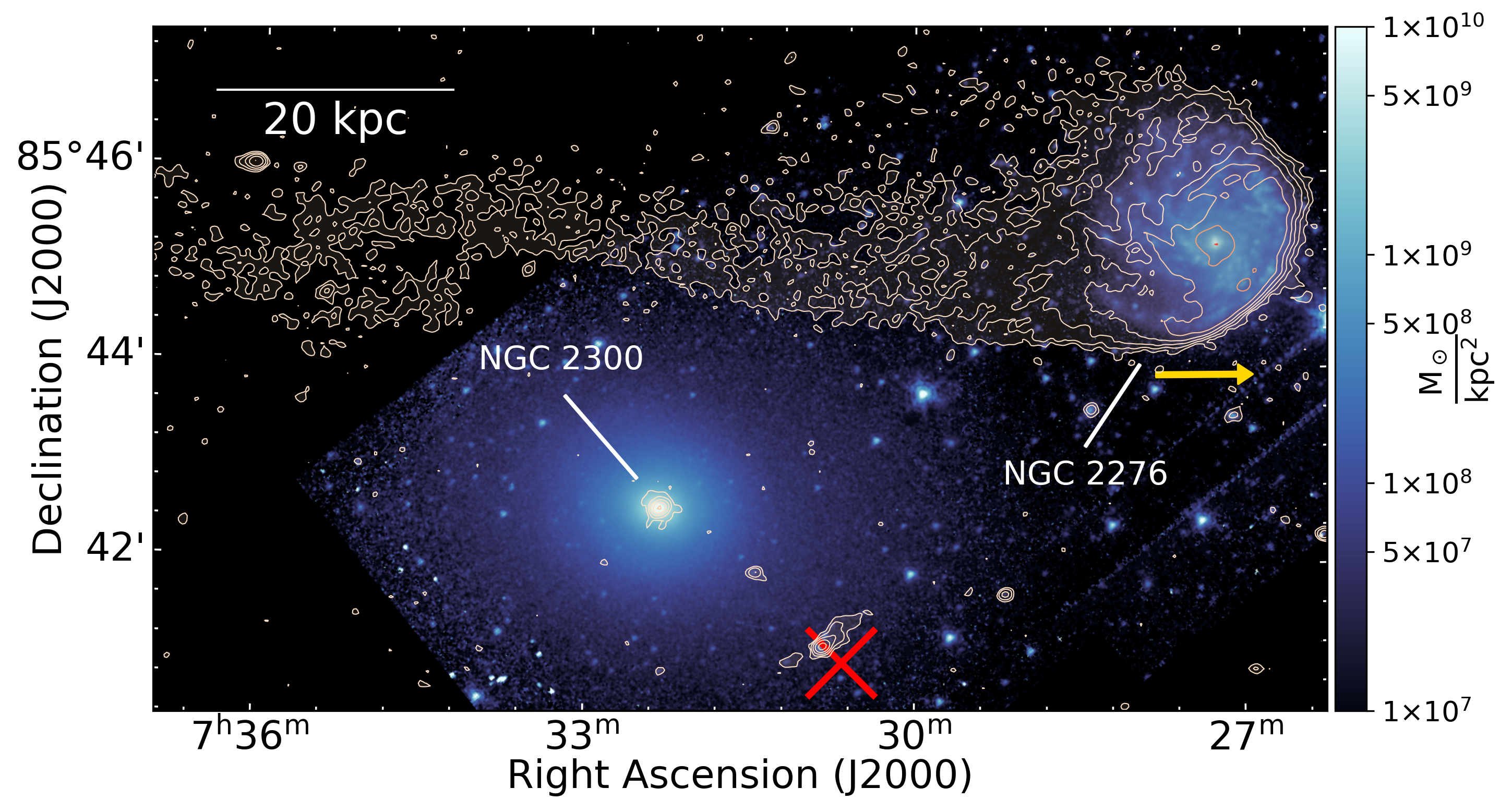}
        \caption{Stellar mass surface density map of the center of NGC 2300 group obtained by combining Spitzer's IRAC 1 and IRAC 2 images using stellar mass relation from \citet{querejeta}. Red cross shows the group X-ray IGM emission center from \citet{mulchaey}. LOFAR contours of the radio continuum emission \citep{roberts} at 144 MHz (3, 6, 12, 24, 48, 96, 192×RMS, RMS = 60$\mu$Jy, resolution = 4.9 $\times$ 4.3 arcsec$^2$) shown overplotted in white. Golden arrow shows the predicted trajectory of NGC 2276, derived from the direction of the radio continuum tail \citep{Souchereau2025}, and the H$\alpha$ kinematics \citep{Haan2014}.}
        \label{fig:N2300_group}
    \end{figure}

    This paper focuses on one of these peculiar systems, NGC 2276, which is a member of the NGC 2300 group (see Fig. \ref{fig:N2300_group}). NGC 2300 group is a nearby galaxy group (z $\approx 0.0067$), located in the Cepheus constellation at equatorial coordinates (J2000.0) $\alpha$ = 07:30:21.35, $\delta$ =  +85:42:00.41. Recent work by \citet{fadda} suggested that this galaxy group consists of 8 members, where the two most massive ones are NGC 2300, and NGC 2276. The detection of diffuse X-ray emission sparked significant interest in this group across X-ray \citep[e.g.][]{hwang}, radio \citep[e.g.][]{davis}, and optical wavelengths \citep[e.g.][]{gruendl, afanasiev, ilina}, since it was the first group in which a diffuse X-ray emitting IGM was detected \citep{mulchaey}. 
    
    The spiral galaxy NGC 2276 (z $\approx 0.0079$, Table \ref{tab:params}) was investigated more thoroughly \citep[e.g.][]{gruendl, davis, wolter15, tomicic18, roberts} due to its interesting morphology firstly described by \citet{arp}.
    The distinct visual features of this galaxy are its asymmetrical shape and the southern spiral arm that does not appear to be tightly wound \citep{zwicky, ngc}, potentially due to interaction with its environment. 
    The galaxy’s leading western side exhibits strong, asymmetric H$\alpha$ emission \citep{gruendl}, associated with intense star formation processes that yield an SFR of $17\pm5\, \mathrm{M_\odot/yr}$ \citep{tomicic18}. 
    Considering a stellar mass of a $\sim5\times10^{10}\, \mathrm{M}_\odot$ (see Table \ref{tab:params}), this high SFR puts NGC 2276 $\sim$ 0.74 dex above the main sequence of star formation \citep{renzini}. 
    A shock-enhanced [\ion{C}{II}] emission has also been observed on the western (leading) side of the galaxy \citep{fadda}, as well as numerous compact X-ray sources \citep{wolter11, wolter15}, where some of them are accompanied by powerful jets \citep[e.g.][]{mezcua}. 
    The southwest edge of the galaxy also showed a sharp edge in both the 20-cm radio continuum, and \ion{H}{I} observations by \cite{davis}.
    The eastern side of the galaxy shows extended X-ray emission \citep{rasmussen}, and a 100-kpc long radio continuum tail \citep{roberts}.
    
    NGC 2276 is moving westward on the celestial sphere through the IGM density of $\sim5\times10^{-27}\mathrm{g\,cm^{-3}}$  \citep{mulchaey, davis96, rasmussen} subjecting its leading side to ram pressure. Its plane of the sky trajectory is indicated by the compressed multi-wavelength morphology on the western side \citep[e.g.][]{gruendl, davis}, and the extended radio continuum emission on the eastern side \citep{roberts}.
    Considering the trajectory of NGC 2276 in the plane of the sky and the projected distance between NGC 2276 and NGC 2300 (see Fig. \ref{fig:N2300_group}), the possibility of a tidal interaction between these two systems has also been suggested.
    Previous works argued that there are evidence of past tidal interactions in the form of: 1) R-band continuum disk truncation in NGC 2276 \citep{gruendl}, 2) non-uniform north-south distribution of the magnetic field \citep{hummel}, 3) the northeastern extension in the I-band continuum of NGC 2300, as well as the presence of stellar shells \citep{forbes}, and 4) the indication of bar-like features in B-V maps of NGC 2300 \citep{ilina}.
    One of the clearest and most convincing signs of whether this galaxy's morphology is mostly influenced by ram pressure would be a relatively undisturbed old stellar disk that extends beyond the gas disk on the western side. This has not yet been investigated in the literature.

    The relative contributions of ram pressure and tidal effects to the morphological and kinematic peculiarities observed within galaxy groups remain a subject of ongoing debate. While ram pressure is a well-established mechanism for driving galaxy evolution in dense cluster environments, its role in the less-extreme conditions of galaxy groups is not yet fully understood. Investigating the influence of ram pressure in this setting is crucial, as it may be a significant driver of evolutionary pathways, particularly for low-mass galaxies, within the environments where the majority of galaxies reside.
    
    This paper is a continuation of the investigations by \citet{tomicic18} and \citet{roberts}, whose goal is to establish the environmental effects on star formation in this system, by using a multiwavelength dataset. Here we focus primarily on Hubble Space Telescope (HST) observations because they deliver high spatial resolution and significant depth in flux, enabling the clear distinction of stellar clusters and spiral arms within the galaxy. To further constrain the stellar population distribution, we complement these data with archival Spitzer observations and Calar Alto IFU spectroscopy.
    
    In Sect. \ref{sec: data}, we present details about the NGC 2300 group, focusing on the galaxies NGC 2300 and NGC 2276. 
    In the same section, we describe the dataset that we use and show the first reduced HST WFC3/UVIS images of NGC 2276. 
    In Sect. \ref{sec: res}, we present our analysis of the spatial distribution of the broad-band light and of the individual spiral features of NGC 2276, and in Sect. \ref{sec: dis} we interpret our findings in terms of environmental effects.
    In Sect. \ref{sec: sum}, we summarize our work on this galaxy. Throughout this paper, we assume a standard $\Lambda$-CDM cosmology model with $\Omega_\mathrm{m}$ = 0.315, $\Omega_\Lambda$ = 0.685, and $H_0$ = 67.4 km s$^{-1}$ Mpc$^{-1}$ \citep{planck}.
    
\section{Observational data and data reduction}
\label{sec: data}

    The observational data used in our study span wavelengths from the near-ultraviolet (NUV; $\approx2710\, \AA$) to the near-infrared (NIR; $\approx4.5\, \mu$m; see Table \ref{tab:filters}). Below, we describe the datasets in detail.
    
    Our analysis focuses on NGC 2276, a member of the NGC 2300 group. Based on the systemic velocity of the group, \citet{fadda} derived a luminosity distance of 28.5 Mpc to the center of the group (see Table \ref{tab:params}). 
    The projected distance between the centers of NGC 2276 and NGC 2300 is $\approx$380$''$, corresponding to $\sim$ 52 kpc at this distance. We use this luminosity distance as the true distance of NGC 2276 throughout this paper, rather than the distance inferred from its redshift alone \citep[35.5 $\pm$ 2.5 Mpc;][]{ackermann}, as the latter is skewed by internal group kinematics \citep{santucho}, as well as local deviations from Hubble flow of the group due to local large scale structure. This choice is also justified by the high peculiar velocity of NGC 2276 compared to other members of the group \citep{tomicic18, fadda}.
    
    The multiwavelength analysis that we present below is based on two assumptions. The first is that the light in different broad-band filters is dominated by stellar populations of different ages (e.g. broadband ultraviolet filters trace mostly younger stellar populations, while broadband infrared filters trace mostly older stellar populations). The second is that one can distinguish between different environmental mechanisms by studying the spatial distribution of stars with different ages, as tidal effects influence both the stellar and gas distribution \citep[e.g.][]{eneev73, gnedin03}, whereas RPS affects primarily gas (and, in turn, the most recent stellar populations) creating asymmetries in the stellar populations \citep[e.g.][]{abadi99, kapferer}.

    \subsection{Hubble Space Telescope Dataset}
        We make use of publicly available data from the HST's Wide Field Camera 3 (WFC3) instrument by \citet{wfc3} (observation ID 15615, P.I. Sell, P., available through the MAST portal). NGC 2276 was observed in separate HST visits from the 31st of July until the 22nd of August 2019 in five broadband UVIS filters, namely F275W, F336W, F438W, F555W, and F814W (details in Table \ref{tab:filters}). The angular resolution (corresponding to the full width at half maximum of the point-spread function; PSF) of these UVIS filters\footnote{\label{foot:wfc3}\url{https://hst-docs.stsci.edu/wfc3ihb/}} is $\approx$0.07$''$ (corresponding to 9.6 pc at the galaxy's luminosity distance), and the field of view (FOV) is $162''\times 162''$ (corresponding to $22\times22$ kpc$^2$).

    \begin{figure*}[]
        \centering
        \includegraphics[width = 0.97\hsize]{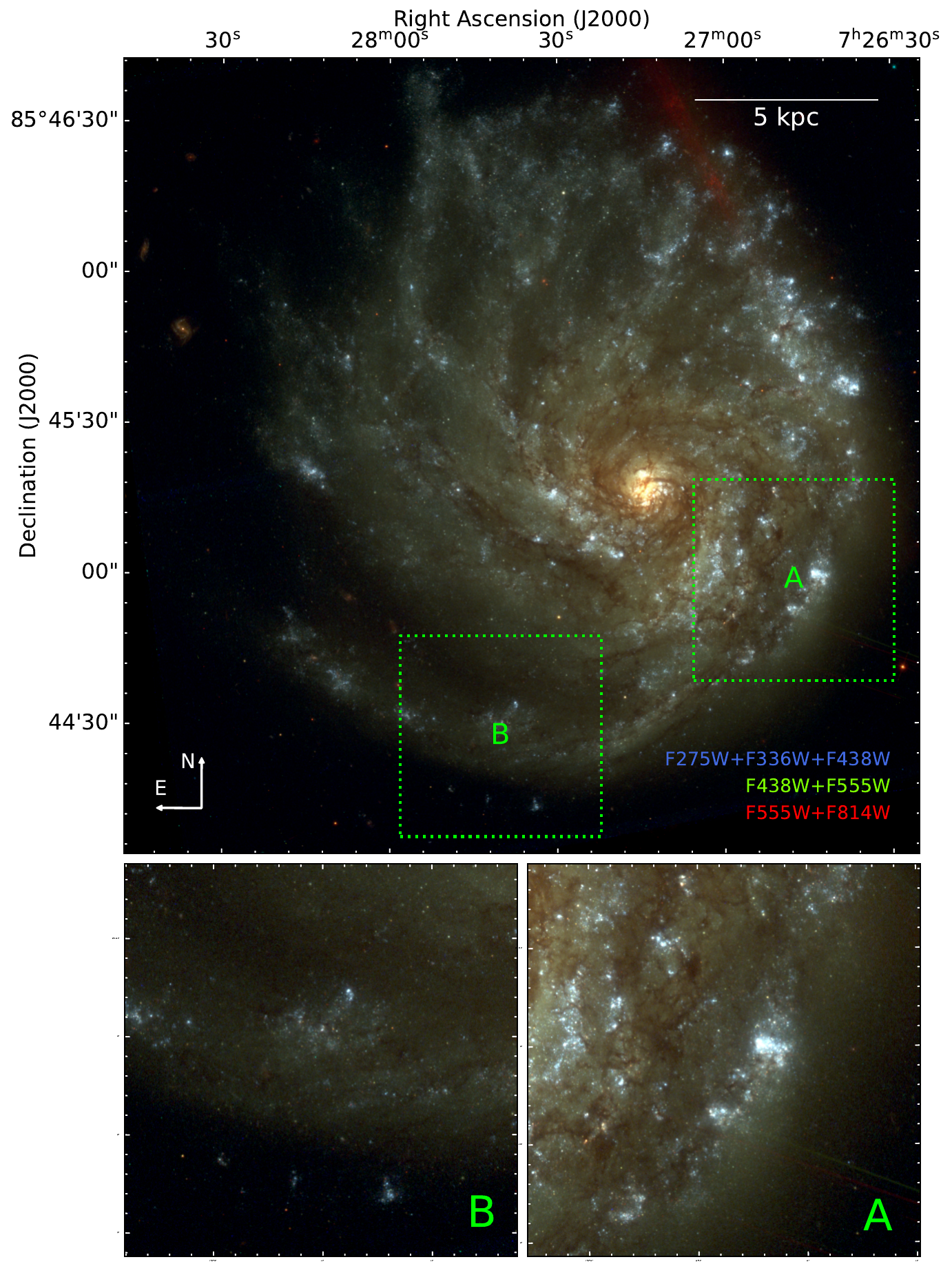}
        \caption{\textit{Top:} color-composite HST image of NGC 2276. The combination of filters used as RGB channels are indicated with corresponding colors on the bottom right. See text for details on the relative weights used for each filter.
        \textit{Bottom Right:} Zoomed in view of the southwestern side of NGC 2276 (Region A) showing disk truncation and the presence of dust.
        \textit{Bottom Left:} Zoomed in view of the southern side of NGC 2276 (Region B) showing isolated clumps of young stars.
        }
        \label{fig:RGB}
    \end{figure*}  

        The images were calibrated, flat-fielded and charge transfer efficiency (CTE) corrected. Each filter has four exposures, each with its own science image, error image, and data quality image. For further data reduction we used \textsc{Astrodrizzle} \cite{astrodrizzle} software. We used the standard \textsc{Astrodrizzle} pipeline to correct the astrometry of individual exposures, stack them, and remove the cosmic rays. The F814W image, and the F555W image in a smaller amount, were contaminated by scattered light from bright sources outside of the detector's FOV \citep{fowler}. Considering that our analysis does not rely critically on the precise photometry on smaller spatial scales, we decided not to reduce these contaminated parts of the images.  
        
        After that, we also subtracted the background level by measuring the median value in 6 apertures (4.5$''$ radius in size) placed in empty regions of the FOV around the galaxy, and applying a tilted-plane background model using \textsc{Astropy's} \citep{astropy} function \texttt{Polynomial2D}. We used the cleaned intensity maps, and divided them with the standard deviation of the background (measured as a median value of the standard deviation from the 6 apertures), to create uncertainty maps. 

        By combining all HST filters, we created a brand new RGB composite map, shown in Fig. \ref{fig:RGB}, such that each color corresponds to each filter with the following relations:
        \begin{displaymath}
            \begin{split}
                \mathrm{B} &= 0.3 \times \mathrm{F275W} + 0.4 \times \mathrm{F336W} + 0.3 \times \mathrm{F438W}, \\
                \mathrm{G} &= 0.5 \times \mathrm{F438W} + 0.5 \times \mathrm{F555W}, \\
                \mathrm{R} &= 0.3 \times \mathrm{F555W} + 0.7 \times \mathrm{F814W}
            \end{split}
        \end{displaymath} 
        F275W and F336W filters had to be rebinned to a 4 times larger pixel size (0.04$''$ to 0.16$''$) to increase their signal-to-noise (S/N) ratio, and we applied this lower resolution for the other HST images as well. The visible red strip in the northwestern part of the RGB image is the effect of scattered light.

    \begin{table}[]
    \caption{General characteristics of instrument filters used in this work.}             
    \label{tab:filters}   
    \centering
    \begin{tabular}{l l l l l l }
    \hline
    \hline
    Filter & $\theta$ [$''$] & $\lambda_\mathrm{eff}$ [$\mu$m] & $\Delta\lambda$ [$\mu$m] & Exp. time & $\mu_\mathrm{lim}$\\
    \hline                                                                        
    F275W       & 0.08         & 0.271                        & 0.0405          & 4$\times$700s     & 22.4     \\
    F336W       & 0.07         & 0.335                        & 0.0512          & 4$\times$700s     & 23.4     \\
    F438W       & 0.07         & 0.433                        & 0.0615          & 4$\times$700s     & 24.0     \\
    F555W       & 0.07         & 0.531                        & 0.1565          & 4$\times$700s     & 25.0     \\
    F814W       & 0.08         & 0.804                        & 0.1565          & 4$\times$700s     & 25.2     \\
    IRAC1       & 1.66         & 3.551                        & 0.75            & 712s              & 23.5     \\
    IRAC2       & 1.72         & 4.493                        & 1.01            & 712s              & 22.5     \\
    \hline
    \end{tabular}
    \tablefoot{$\lambda_\mathrm{eff}$ is the effective wavelength of a filter, $\Delta\lambda$ is the passband rectangular width, $\theta$ is the full width at half maximum (FWHM) of the Gaussian point spread function (PSF) for the corresponding filter, and $\mu_\mathrm{lim}$ is the 1-$\sigma$ limiting surface brightness in mag/arcsec$^2$. }
    \end{table}

    \subsection{Ancillary data}
    
        In addition to the newly reduced HST images, we used NIR archival data from the Spitzer's Infrared Array Camera (IRAC, observation ID 20140, P.I. Zezas, A.). Using NASA/IPAC's Infrared Science Archive (IRSA), we retrieved Spitzer's IRAC images from channels 1 and 2 (referred to as IRAC1 and IRAC2), centered on  3.6$\,\mu$m and 4.5$\,\mu$m, with angular resolution of 1.66$''$ and 1.72$''$, respectively. These images cover a FOV of $\approx$1080$''\times$410$''$ centered on NGC 2276. In IRAC1 and IRAC2 images we subtracted the background level using 9 apertures of 7.5$''$ radius in size centered on locations devoid of visible sources. This dataset is used to trace older stellar populations, mapping the stellar mass distribution (shown in Fig. \ref{fig:N2300_group}), and mapping the size of the older galactic disk of NGC 2276.

        Furthermore, we used optical IFU data from the PMAS/PPaK instrument at the Calar Alto 3.5m telescope (Spain). The reduced spectral cube has a spectral resolution of $R\approx$ 1000, covers a wavelength range of $3700-7010\AA$, and has an angular resolution of 2.7$''$. From the reduced data cube, we extracted the H$\alpha$ moment-0 and moment-1 maps corrected for dust attenuation (A$_V$) using the Balmer decrement, assuming an intrinsic ratio of H$\alpha$/H$\beta$ = 2.863 for a gas temperature $T_e$$\sim10^4$K \citep{osterbrock} and the dust extinction law of \citet{ccm89}. Details of data reduction and emission line extraction are described in \citet{tomicic18}.
        
\section{Results}
\label{sec: res}

   Fig. \ref{fig:RGB} shows that NGC 2276 is a peculiar galaxy with an asymmetric distribution of stellar populations. The southwestern side of the galaxy (region A from Fig. \ref{fig:RGB}), which is on the front of the impact with the IGM gas, is dominated by clumps of young OB stars. 
   There is a visible compressed morphology of the young stellar disk on the leading side of the galaxy. The trailing side of the disk also shows that older and younger stellar populations are extended to different radii, but the effect is less prominent than on the leading side.
   
   There is also an abundance of flocculent-looking dust surrounding the galactic center, and the dust is slightly obscuring the young stellar populations. On smaller scales, there are no visible elongated linear or head-tail dust features as observed in some galaxies experiencing active ram pressure stripping \citep[e.g.][]{abramson14, kenney15}, possibly because their formation requires even stronger ram pressure.
   
   On the southern side of the galaxy there are some clumps of young stars that are farther away from the edge of the young stellar disk, and could possibly be outside of the galactic disk ($\sim 9$ kpc away from the center), as seen in Region B in Fig. \ref{fig:RGB}. 
   Similar features have already been observed in RPS galaxies, and were found to be the evidence of extra-planar SFR induced by the ram pressure \citep[e.g.][]{kenney1999, giunchi}. 
   Considering that the galaxy is viewed face-on, it is difficult to determine whether these clumps were formed from extra-planar gas. 
   Although it is not certain that ram pressure is responsible for the formation of these clumps, since they lack the distinct head-tail shape of the \textit{fireball} model, it remains a plausible cause. 
   The absence of this characteristic morphology might be due to requirement of even stronger ram pressure for their formation, but it remains an open question for future studies.
   The spiral features on the northeastern side of the galaxy seem to show unwinding effects, which we investigate in Sect. \ref{ssec: unwinding}.

    \begin{figure}[t!]
        \centering
        \includegraphics[width = \hsize]{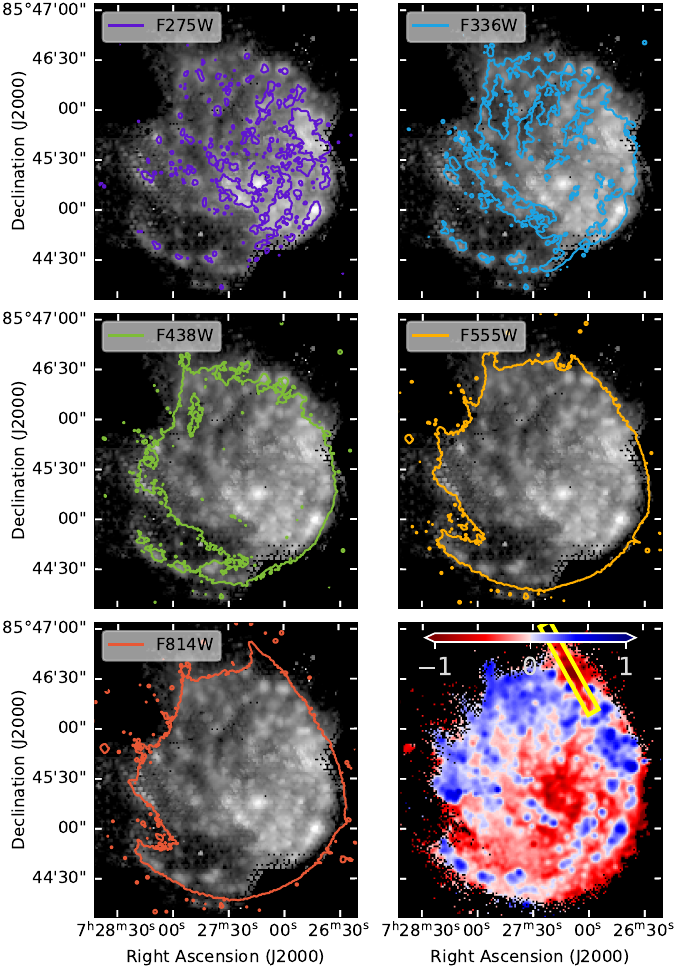}
        \caption{3-$\sigma$ S/N level contours for different HST filters of NGC 2276. The background grayscale image shown in each panel is the H$\alpha$ moment-0 map extracted from the Calar Alto IFU data. The bottom right panel is the color image, expressed as $\log_{10}\left( \mathrm{F275W/F814W}\right) $, smoothed and reprojected to the same resolution as H$\alpha$ map. Yellow rectangle shows approximately the region affected by the scattered light anomaly \citep{fowler}.}
        \label{fig:cont}
    \end{figure} 

   In this section, we present the newly determined morphological properties of NGC 2276 and discuss how the ram pressure and tidal forces had an impact on the morphology of this galaxy.

    \subsection{Distribution of stellar populations}  

        With the RGB composite image, we have a visual representation of how various stellar populations are distributed across the disk of NGC 2276. In this subsection, we will use uncertainty maps of HST WFC3/UVIS filters to show variations in the morphology of different stellar populations with a more quantitative approach. 
        
        In order to see this, we created contours of areas where S/N $\geqslant 3$ in each filter. The results are shown in Fig. \ref{fig:cont}, plotted on top of the H$\alpha$ map. Each contour represents a single HST filter, which is representative for stellar populations of different ages, with blue filters tracing younger stars and red filters tracing older populations.  
        
        As shown in Fig. \ref{fig:cont}, the youngest stellar populations, represented by the F275W and F336W filters, are not evenly distributed across the galaxy. Their distribution is patchy and asymmetrically distributed. The regions exhibiting the highest intensity of the ionized gas, traced by H$\alpha$ emission, coincide with the clumpy distribution of the youngest stellar populations identified in HST filter imaging, showing that UV filters are sensitive to the location of SF regions. SF regions seem to be distributed in smaller clumps on the trailing side, and in larger ones on the leading side of the galaxy.

        The redder filters F555W and F814W show that the older stars are more symmetrically distributed across the galactic disk. What is also visible from Fig. \ref{fig:cont} is that the older stellar populations on the leading side are spread farther away from the center of the galaxy than the younger ones, implying that the leading side is affected by ram pressure. This was partially visible in the color composite image from Fig. \ref{fig:RGB} with the redder colors spreading farther away than the bluer ones.

        As the NUV bands are less sensitive than optical bands (Table \ref{tab:filters}) and do not have the same resolution as the H$\alpha$ map, so they might not encompass all the star forming regions detected with H$\alpha$. That is why we decided to smooth and reproject the F275W and F814W maps to the resolution of the H$\alpha$ map and create a color map that we show in the bottom right panel of Fig. \ref{fig:cont}. This highlights better a feature that was partially visible already in the contours, that is, the northeastern part of the galaxy is dominated by blue color (younger stellar populations), while the southwestern region is dominated by the red color (older stellar populations), hinting towards the truncation of the gas disk.

    \subsection{Determining the location of the center}

    \begin{figure}[t!]
        \centering
        \includegraphics[width = \hsize]{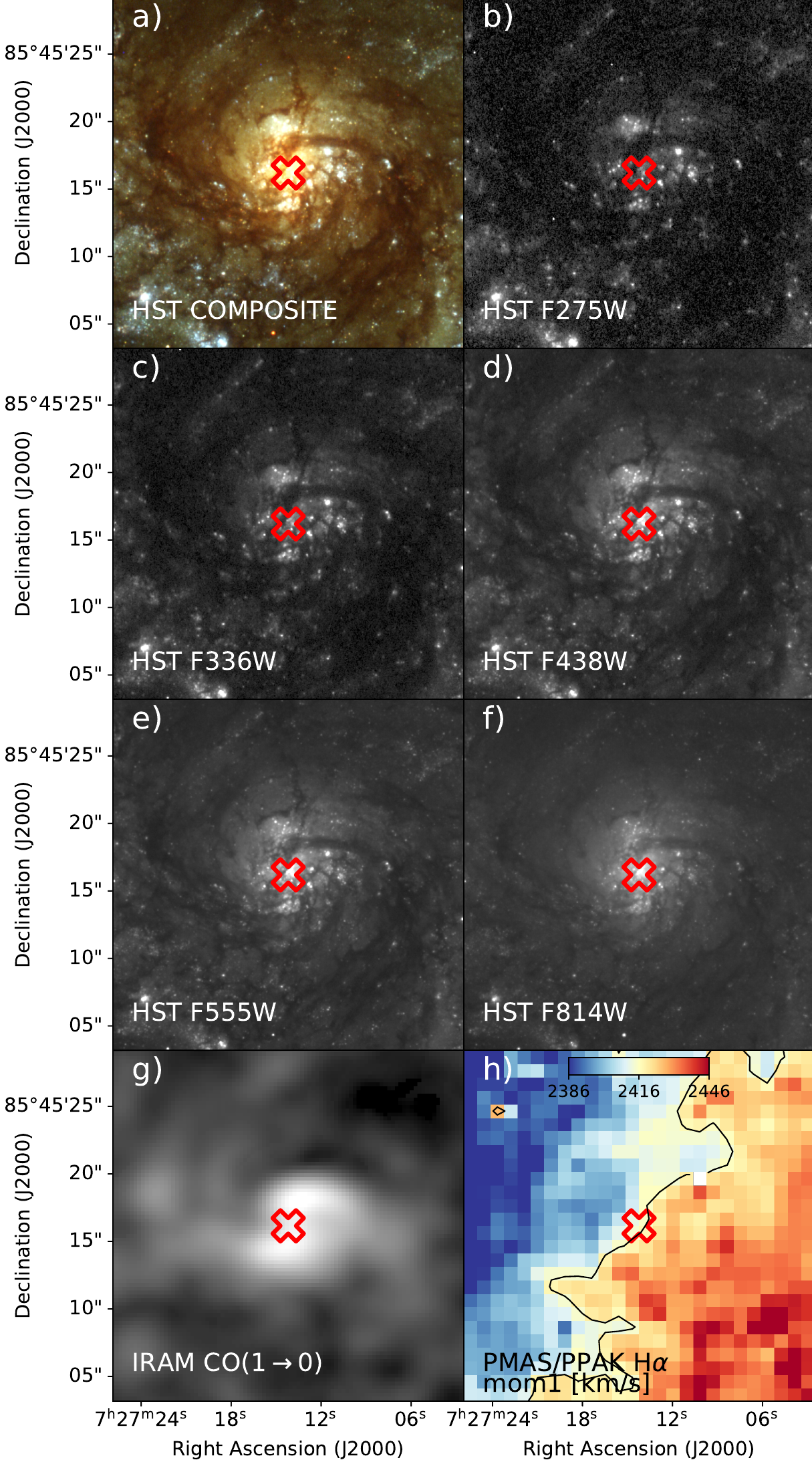}
        \caption{Location of the NGC 2276 center, shown in different tracers. Red cross marks the coordinates of the center. Panel a shows the RGB composite from Fig. \ref{fig:RGB}, panels b-f show HST UVIS/WFC3 filters from NUV to NIR. Panel g shows CO map from the IRAM NOEMA dataset, and in the panel h, the velocity map from PMAS/PPAK map of H$\alpha$ is shown \citep{tomicic18}, along with the black contour depicting the systemic velocity of 2416 km/s.}
        \label{fig:gal_cen}
    \end{figure}

        To better measure galactocentric radii across the NGC 2276's disk, we had to accurately determine the location of its center. 
        In Fig. \ref{fig:gal_cen} we show intensity maps from all five of the HST filters, zoomed in on the galactic center. 

        We searched for the pixel with peak surface brightness intensity on the HST F814W map around the previously determined galaxy center location, which was based on the crossing of the spiral arms of molecular gas. The molecular gas was traced by $^{12}$CO($J=1\rightarrow0$) emission observed with IRAM/NOEMA \citep{tomicic18}. This pixel is enclosed within the aperture the size of the H$\alpha$ moment-1 map resolution (2.7$''$) from the kinematic center of the galaxy. Coordinates $\alpha$ = 7:27:14.2 and $\delta$ = +85:45:16.2 are depicting the galactic center with an uncertainty of the H$\alpha$ map resolution.
        The location of the center is shown in Fig. \ref{fig:gal_cen}, along with the CO and H$\alpha$ velocity map of the center (panels g and h), and the RGB image created from the HST filters (panel a). We can see that the highest intensity of the CO map does not coincide with the highest intensity of the HST maps, but the CO show some spatial correlation with the dust lanes, which are visible in the RGB composite. This high dust content in the center is also hinted from color diagram (Fig. \ref{fig:cont}) where there is a visible increase in the red colors around the galactic center.

    \subsection{Morphology of the stellar disk}

        \begin{figure}[h!]
            \centering
            \includegraphics[width=\hsize]{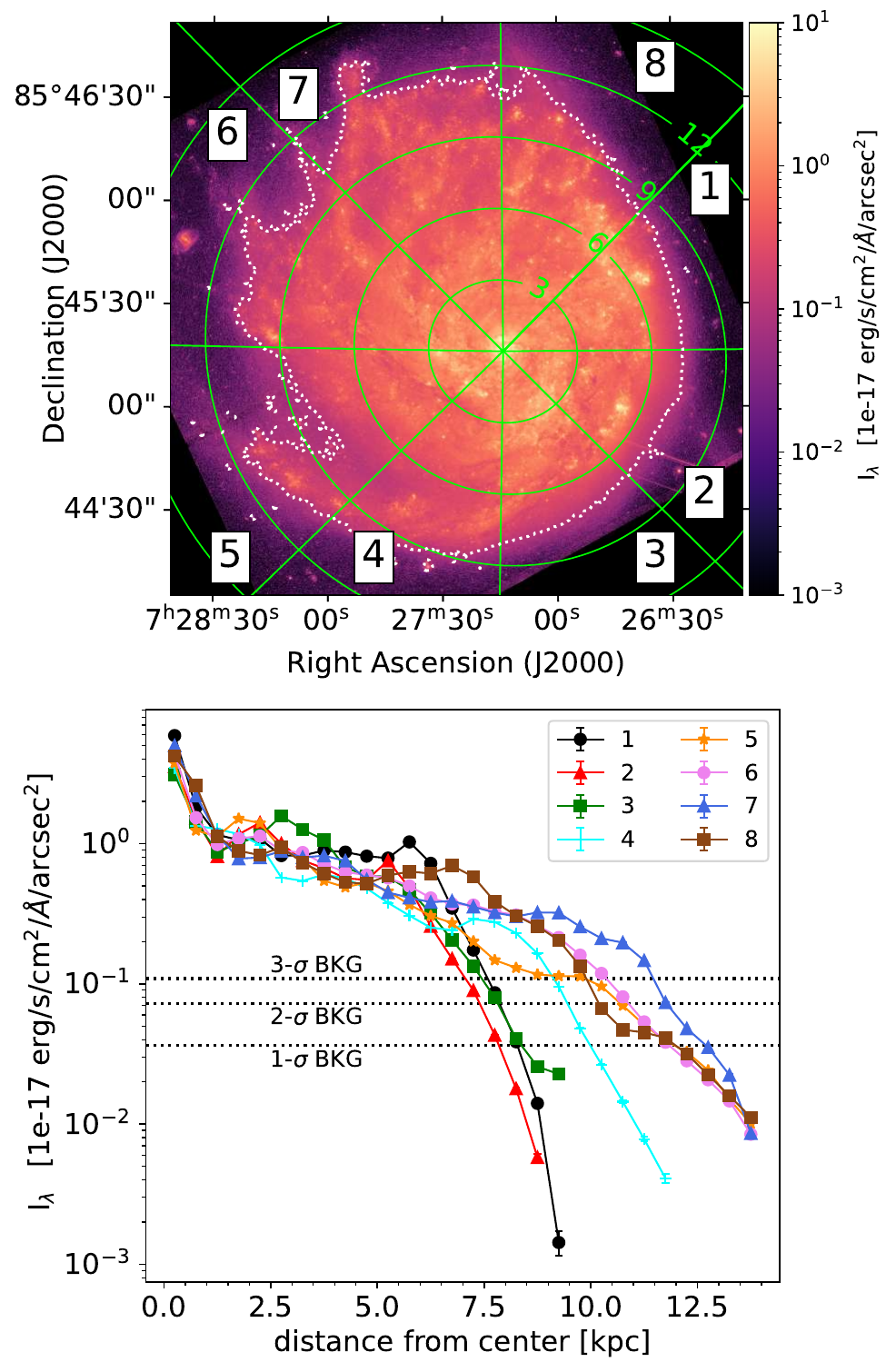}
            \caption{Intensity profiles of NGC 2276 as a function of galactocentric radius. \textit{Top:} F555W image divided in 8 sectors with constant azimuthal separation of 45°. The concentric ellipses show regions with constant  galactocentric distance of 3, 6, 9 and 12 kpc. White dotted line represents 3-$\sigma$ background noise level contour.  \textit{Bottom:} Surface brightness radial profile in the F555W band for each of the 8 sectors in bin size of 0.5 kpc. Three dotted lines represent 1,2, and 3-$\sigma$ background noise level. All data points have S/N above 5.}
            \label{fig:rad_dist_555}
        \end{figure}

        We now quantify both the disk asymmetry and the spatial variation of the stellar population across the galactic disk. We will also determine the size of the galactic disk for both the older and younger stellar populations. 

        In Fig. \ref{fig:rad_dist_555} we show intensity profile as a function of galactocentric radius for the F555W filter across 8 sectors, in order to show the extent of stellar disk asymmetry. The sectors (shown in upper panel of the figure) are equally spaced in 45° angle, and the galactocentric radius is corrected for inclination effects. The intensity values in the bottom panel represent the median intensity in radial bins with size of 0.5 kpc. The combination of values of 45° and 0.5 kpc were chosen because they provided the optimal signal-to-noise ratio compared to the number of pixels per bin.

        \begin{figure}[h!]
            \centering
            \includegraphics[width=\hsize]{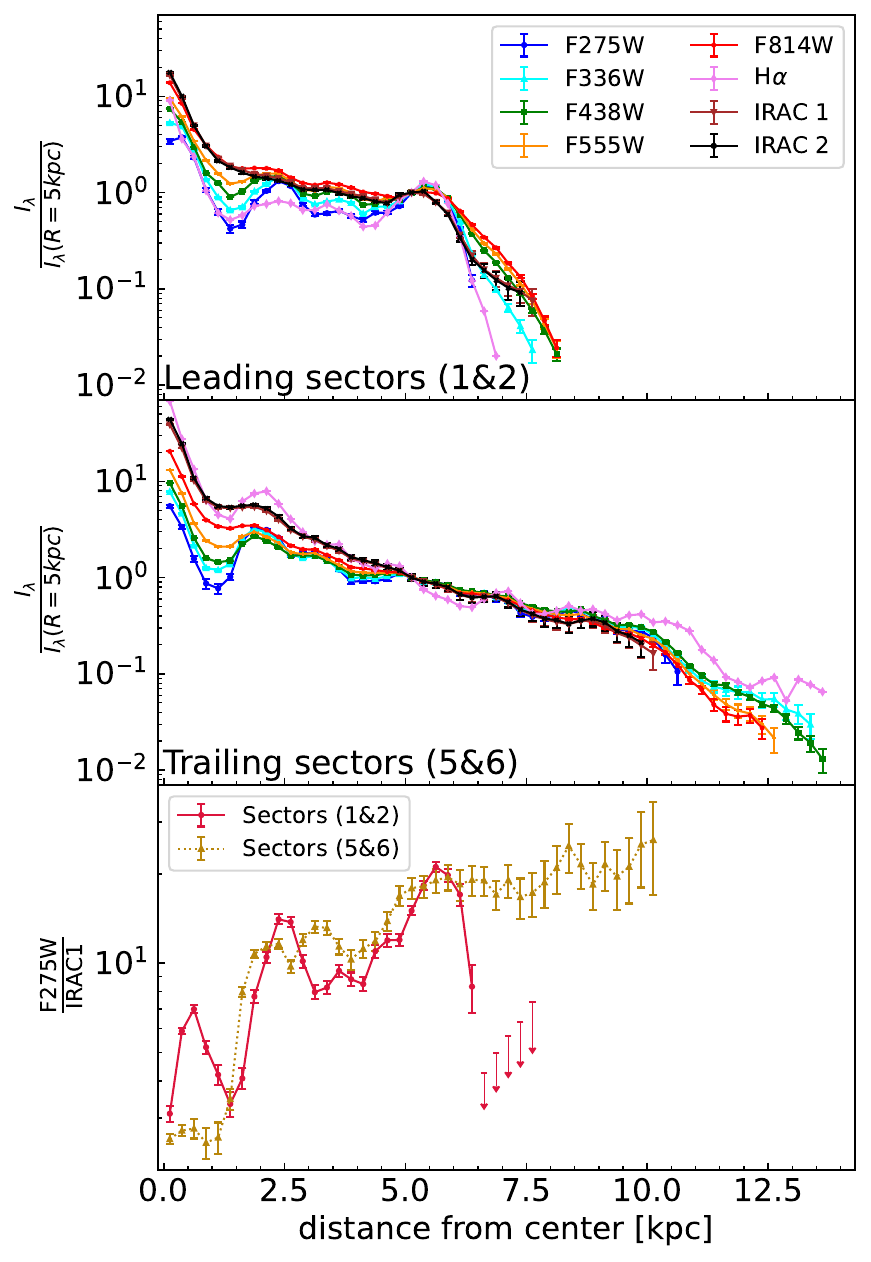}
            \caption{Intensity profiles of leading and trailing sides of NGC 2276 as a function of galactocentric radius. All data points have S/N above 3. \textit{Top:} Radial profiles for the leading side (sectors 1 and 2) for all HST filters, H$\alpha$, and IRAC 1 and IRAC 2. \textit{Middle:} Radial profiles for the trailing side (sectors 5 and 6) for all HST filters, H$\alpha$, and IRAC 1 and IRAC 2.
            \textit{Bottom:} color-radial profile of the leading and trailing side expressed as a ratio of F275W and IRAC 1.
              }
            \label{fig:rad_dist}
        \end{figure}

        The bottom panel of Fig. \ref{fig:rad_dist_555} shows how the azimuthally-averaged intensity in the F555W filter varies as a function of the galactocentric distance in each sector. We determine the extent of the galactic disk in different sectors by calculating the radius that encloses 99 \% of the flux. 99\% of the flux was chosen as a threshold because it encloses all the flux originating from the galaxy up to 2-$\sigma$ integrated S/N inside the bin. On the western side (sectors 1-2, where the leading side is), the radial extent is up to 8 kpc, and on the trailing side (sectors 5-6) it is up to 12.5 kpc. In a recent work by \cite{roberts}, it was shown that this galaxy has a radio tail on the eastern side. These differences in the disk length might suggest, along with the information of the radio tail on the eastern side, the presence of an optical tail (see Fig. \ref{fig:rad_dist_555}).

        \begin{figure}[h!]
            \centering
            \includegraphics[width=\hsize]{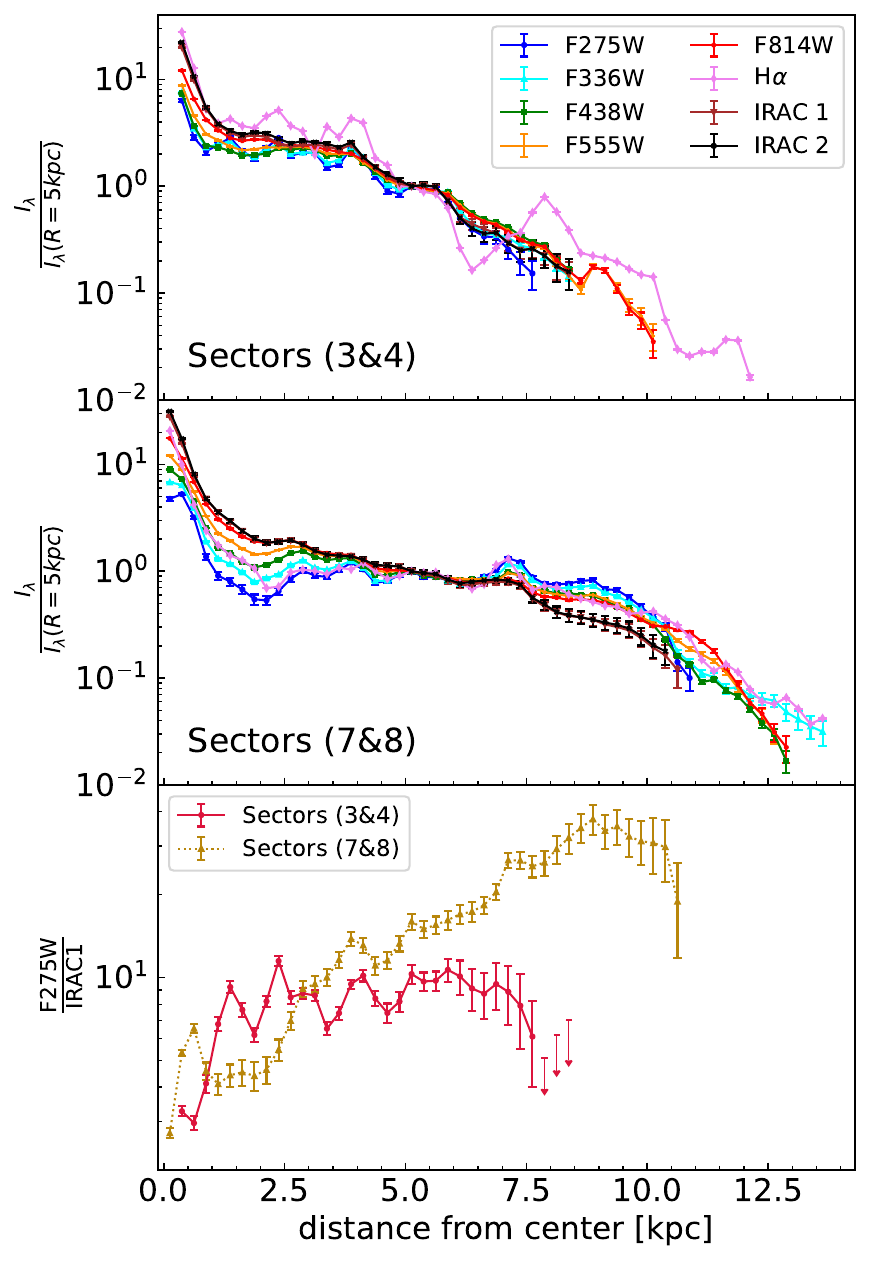}
            \caption{Intensity profiles for the rest of the sectors. All data points have S/N above 3. \textit{Top:} Radial profiles for the side that rotates into the IGM wind (sectors 3 and 4) for all HST filters, H$\alpha$, and IRAC 1 and IRAC 2. \textit{Middle:} Radial profiles for the side which rotates with the IGM wind (sectors 7 and 8) for all HST filters, H$\alpha$, and IRAC 1 and IRAC 2.
            \textit{Bottom:} color-radial profile for two sides of the galaxy shown in upper panels expressed as a ratio of F275W and IRAC 1.
              }
            \label{fig:rad_dist_48}
        \end{figure}

        In Fig. \ref{fig:rad_dist}, we studied how the intensity profile in the HST images changes when we move from the leading side (sectors 1 and 2) to the trailing side (sectors 5 and 6) for each filter, compared to the H$\alpha$ emission , and the infrared emission traced with IRAC 1 and IRAC 2. However, we must take into account that in the case of H$\alpha$ emission, we are limited by the FOV of observations. When using multiple sectors, for a given galactocentric bin, we take the median of the intensity values throughout all sectors used. In this analysis we smoothed and reprojected all our maps to our lowest resolution map (H$\alpha$, 2.7$''$ FWHM PSF, 1$''$ pixel scale), and decreased the radial bin size to 0.25 kpc.
        The various panels in Fig. \ref{fig:rad_dist} indicate the two main key features of the light distribution of NGC 2276. 
        First, there is a striking difference between the surface brightness profiles in the leading side (sectors 1 and 2) and in the trailing side (sectors 5 and 6), with the former showing a break at R$\approx$6 kpc \citep[comparable to the location of the break found by][when adjusted for the newly determined distance to the group by \citealt{fadda}]{gruendl}, and the latter proceeding virtually undisturbed up to R$\approx$12 kpc and beyond.
        The distribution of H$\alpha$ emission closely coincides with the young stellar populations visible in the NUV HST filters (Fig. \ref{fig:cont}), while IRAC 1 and 2 filters show similar extent as reddest HST filter, but with a different slope.
        Second, in a given sector of the disk, different stellar populations exhibit a different profile: specifically, the younger stars seem to be distributed closer to the galaxy center than the older stars on the leading side, while vice versa happens on the trailing side. These two features are well summarized in the bottom panel of Fig. \ref{fig:rad_dist}, which clearly shows that there is a steeper downturn towards the red color on the leading side at $\sim$ 6 kpc, while the trailing side becomes bluer with increasing radius. Fig. \ref{fig:rad_dist} indicates that both the young and the old stellar populations are asymmetrically distributed with respect to the center.

        The fact that the asymmetry is visible in all filters (i.e., all stellar populations), and that there is an increase towards the red colors on the leading side, while the trailing side is becoming increasingly blue, supports the hypothesis of an RPS origin for the features discussed. In the next section we discuss of both ram pressure and tidal interaction being at work in this peculiar galaxy.        

        In Fig. \ref{fig:rad_dist_48}, we analyze the four remaining sectors (The top panel showing the sectors 3 and 4, and the central one showing sectors 7 and 8). Based on the H$\alpha$ velocity map, the assumption of trailing spiral arms, and the features of the radio continuum tail, we determined that sectors 3 and 4 rotate into the IGM wind, while sectors 7 and 8 rotate with it \citep{Souchereau2025}. 
        Radial profiles of sectors 3 and 4 also show an asymmetric distribution compared to sectors 7 and 8.  Sectors 3 and 4 are $\sim 2$ kpc closer to the galactic center compared to sectors 7 and 8, which is most probably an effect of RPS, due to them rotating into the IGM, while 7 and 8 rotate with it. Additionally, when we observe the colors, shown in the bottom panel of Fig. \ref{fig:rad_dist_48}, we see that sectors 3 and 4 show an enhancement towards red colors, while the sectors 7 and 8 show an enhancement towards the blue colors. Similar thing could be observed from the color map shown in Fig. \ref{fig:cont}.  
        These results will be discussed further in Sect. \ref{sec: dis}.

    \subsection{Measuring the pitch angles of spiral arms}
    \label{ssec: unwinding}

     \begin{figure}[h!]
        \centering
        \includegraphics[width=\hsize]{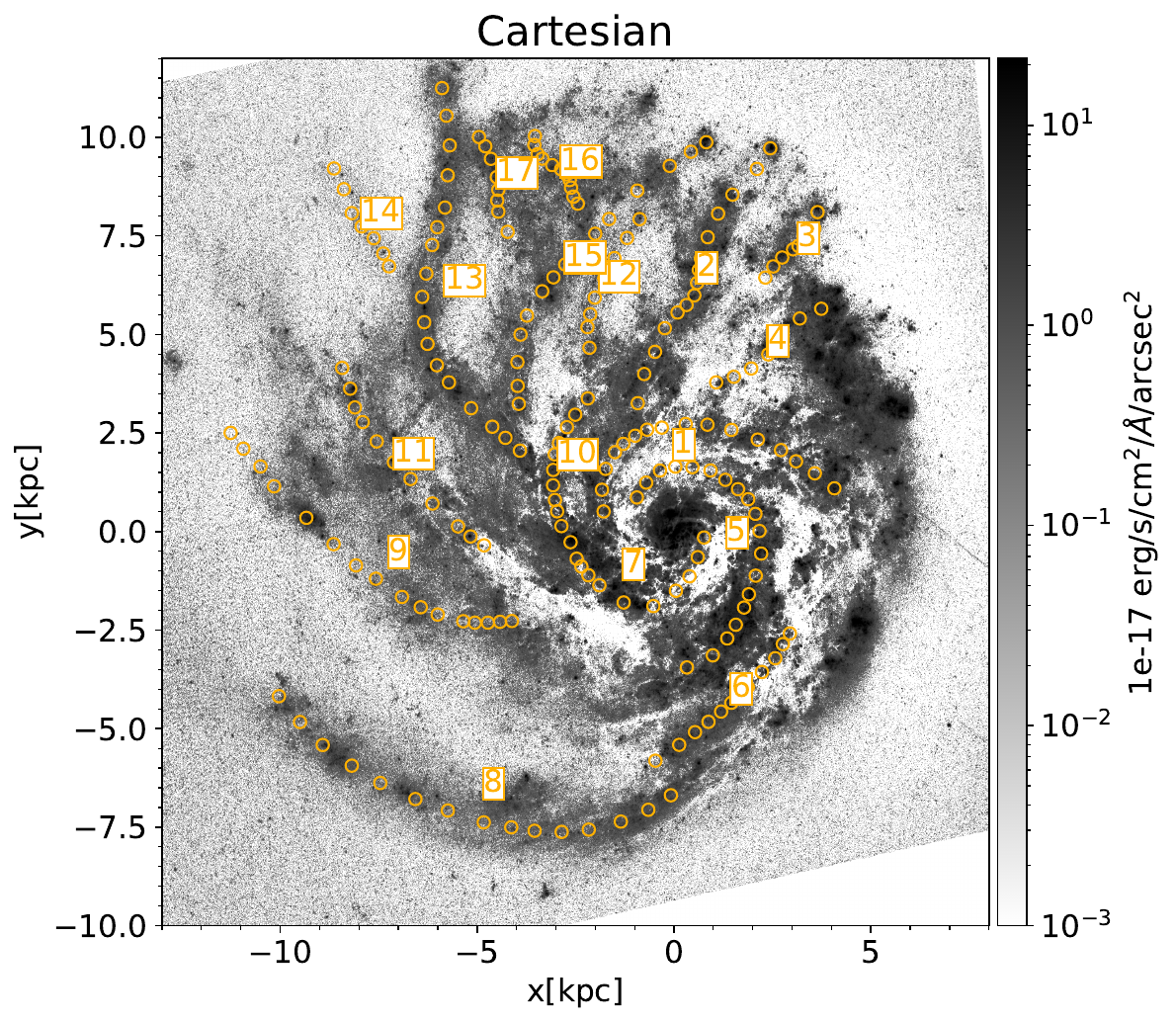}
        
        \includegraphics[width=\hsize]{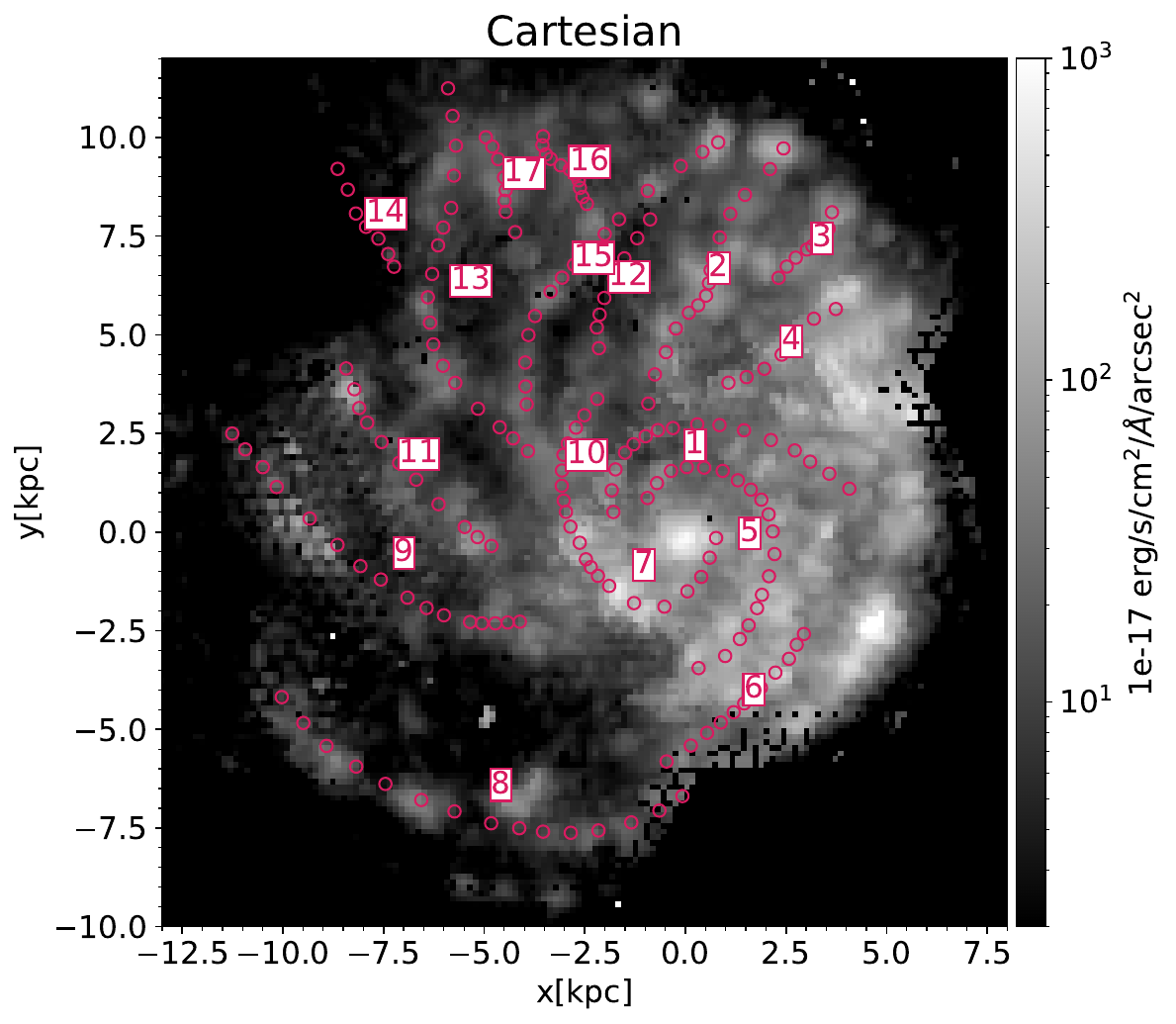}

        \caption{\textit{Top:} Locations of 17 overdensity features found in NGC 2276 shown over-plotted on the deprojected and sharpened WFC3/UVIS F438W filter.
         \textit{Bottom:} Locations of 17 overdensity features found in NGC 2276 shown over-plotted on the deprojected and sharpened H$\alpha$ map.
          }
        \label{fig:sp_arms}
    \end{figure}    

    \begin{figure*}[h!]
        \centering
        \includegraphics[width = \hsize]{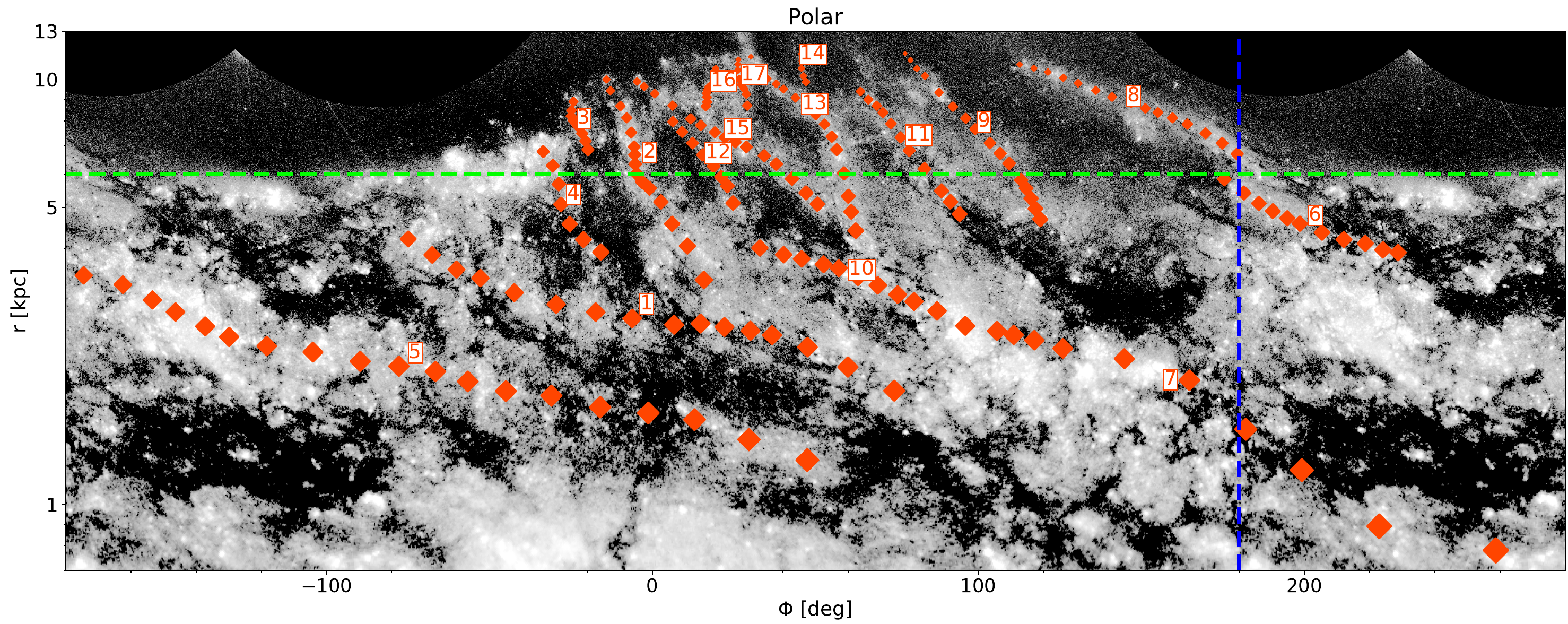}
        \caption{Spiral arm locations shown on the deprojected and sharpened WFC3/UVIS F438W filter in polar coordinates. Radial distance is shown on a logarithmic scale to represent overdensity features as straight lines. Lime green horizontal line represents 2 R$_\mathrm{eff}$. Blue vertical line shows the azimuthal angle of 180°.
          }
        \label{fig:sp_arms_F438W_unwound}
    \end{figure*}

    \begin{figure}[h!]
        \centering
        \includegraphics[width=\hsize]{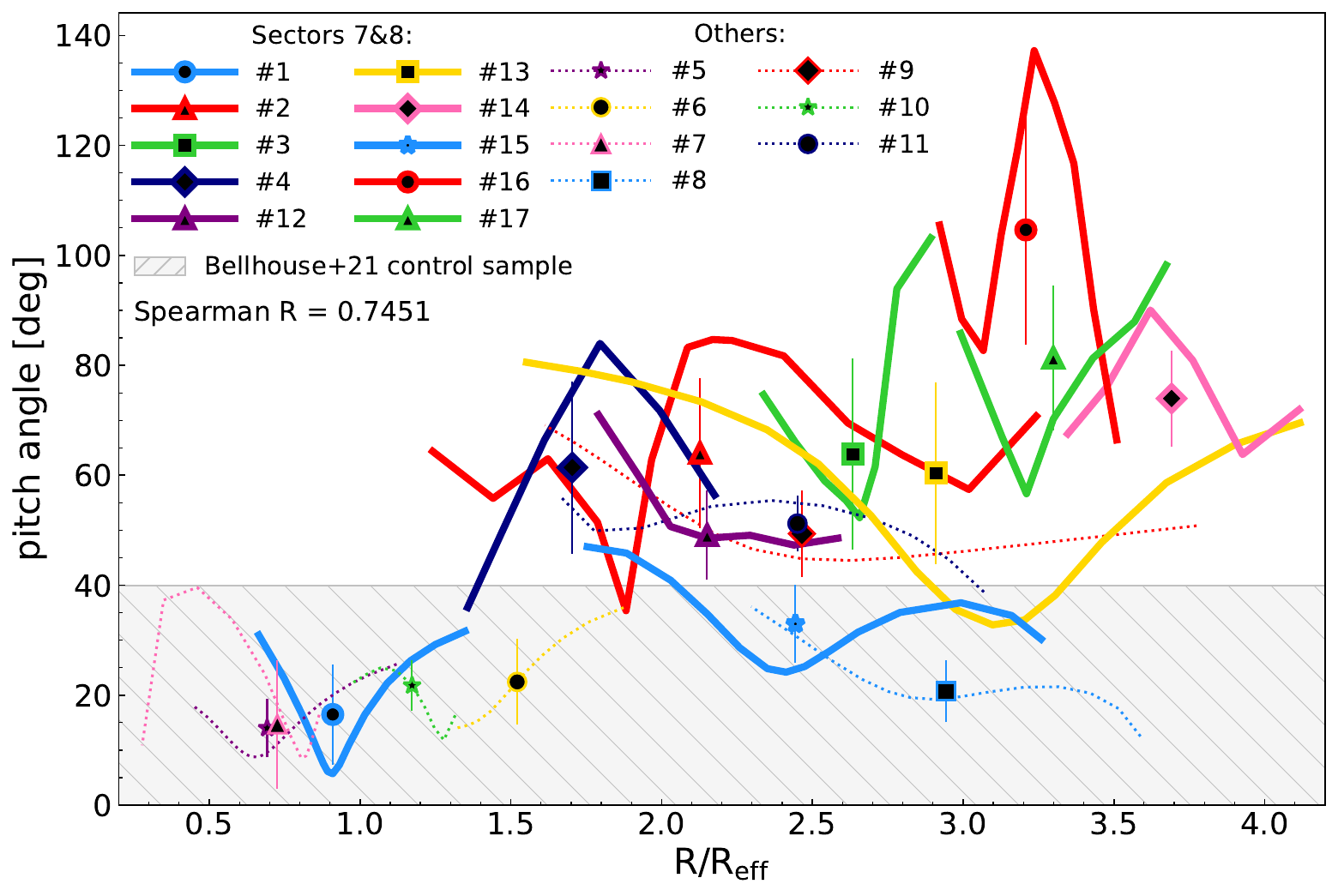}
        \caption{Pitch angles for the 17 overdensity features found in NGC 2276. Bold and solid lines represent overdensity features in sectors 7\&8), while the thin and dotted lines represent all the others. The points show the median value of the pitch angle in each of the overdensity features, and the error bars represent the standard deviation. Shaded area shows the highest pitch angle values for the sample of undisturbed galaxies from \citet{bellhouse}.
          }
        \label{fig:pitch_ang}
    \end{figure}

    One of the possible visible effects of the ram pressure is the unwinding of the spiral arms, where the material in the spiral arms is being pushed away from the force of the IGM. This is caused by the combined effect of galaxy's rotation and the direction of the IGM wind \citep[e.g.][]{bellhouse, Machado2025}. To quantify this effect in NGC 2276, we first need to trace the locations of the spiral arms and then measure their pitch angles.

    To identify the locations of the spiral arms in NGC 2276, we used the H$\alpha$ emission map, IRAC1 map, and F438W map. 
    H$\alpha$ was used instead of the HST UV maps to trace young stars, IRAC1 for older stars, and F438W map was used in the optical wavelengths considering it was the only HST optical map not contaminated by the scattered light. 
    Firstly, we deprojected each of these three images to a face-on orientation, and then applied the unsharp mask filter to enhance the edges of the spiral arms. 
    The unsharp mask filter was generated by convolving the deprojected map with a Gaussian kernel with a standard deviation $\sigma$ of 2 kpc, by knowing that the approximate width of the spiral arms is 2 kpc \citep{block}. 
    Subsequently, the difference between the deprojected map and the convolved map was computed, multiplied by a factor of two, and added to the original deprojected image. Different multiplication factors were tested, and based on visual assessment, a factor of 2 was determined to be the most suitable.
    Afterwards, we compared all three tracers to find where the locations of the overdensity features coincide. 
    We identified them by visual inspection and found 17 potential overdensity features, which are represented by a set of points in Fig. \ref{fig:sp_arms} overlaid on the sharpened F438W image (top panel) and H$\alpha$ emission-line map (bottom panel). 
    Considering the possible effects of the ram pressure and the rotation of the galaxy itself, the leading side of the galaxy contains a potential intertwining of the spiral arms. 
    Therefore, we decided not to trace any of the overdensity features directly on the leading side. Fig. \ref{fig:sp_arms} provides indication for the unwinding of the spiral structure in NGC 2276. We now proceed to quantify this effect.

    To derive the pitch angles of the overdensity features, we first projected our maps onto cylindrical coordinates, as illustrated in Fig. \ref{fig:sp_arms_F438W_unwound}. 
    The lime green horizontal dashed line represents 2 R$_\mathrm{eff}$ \footnote{We calculated R$_\mathrm{eff}$ by analyzing the azimuthally averaged surface brightness profile of F814W image on the leading side.}. 
    The vertical blue dashed line represents the azimuthal angle of 180°, and 0 degrees is measured from the center towards the north. 
    The pitch angle $\theta$ is represented with the relation describing the radius of the spiral arm: \begin{equation} 
                \ln R = \ln R_0 + (\phi-\phi_0)\cdot\tan\theta,
        \label{eqn:pitch}
    \end{equation} 
    where R is the distance measured from the center of the galaxy, $\phi$ is the azimuthal angle, and $R_0$ and $\phi_0$ are normalization constants. When converted to differential form this relation becomes:
    \begin{equation}
        \tan\theta = \frac{\Delta\ln R}{\Delta \phi}.
        \label{eqn:pitch_delta}
    \end{equation}
    We tested the hypotesis of the unwinding spiral arms by calculating how the pitch angle changes from one point to another in each of the overdensity features by tracing the pitch angle radial profile in each of the overdensity features using Eq. \ref{eqn:pitch_delta}. 
    Before applying Eq. \ref{eqn:pitch_delta} to our set of points, we smooth the data using the \texttt{Univariate spline} technique based on an algorithm by \citet{spline}, as implemented by \textsc{scipy} Python package \citep{scipy}. 
    We adopted the parameters\footnote{$s$ is the smoothing factor; and $k$ is the degree of smoothing spline.} of $s=0.1$ and $k=5$. 
    This function was used to smooth the $\ln$ R($\phi$) profiles of the various spiral features in order to avoid discontinuities in its first derivative. 
    We verified by visual inspection that the spline function provides a very good description for all the profiles observed.
    
    The results of our calculations are shown in Fig. \ref{fig:pitch_ang}. 
    The spline shows how the pitch angle changes its value with distance, and the points represent the median value of the pitch angle for that overdensity feature. 
    The figure also shows the shaded area that traces the maximum pitch angle value for undisturbed galaxies from \citet{bellhouse}.
    Our results indicate that the pitch angle globally increases throughout the disk radially from the galactic center, as was also shown in \cite{bellhouse} on a sample of galaxies affected with ram pressure. We also find indication that this increase starts beyond $\approx$ 2 R$_\mathrm{eff}$. From Fig. \ref{fig:sp_arms} and Fig. \ref{fig:pitch_ang} we also observe that the pitch angles are higher on the side that is rotating with the IGM wind (sectors 7 and 8, e.g. overdensity features \#2 and \#3), than on the side that rotates into the IGM wind (sectors 3 and 4, e.g. overdensity features \#6 and \#8). Similar effect was observed in simulated galaxies undergoing edge-on stripping in \citet{bellhouse}.

\section{Discussion}
\label{sec: dis}

So far, the research in the literature was inconclusive on whether ram pressure or tidal interaction is the main cause of the peculiar morphology in NGC 2276. In this section, we will try to discern which effect is prevalent by taking into account all previous findings on ram pressure and tidal interaction on NGC 2276, as well as on other galaxies.

\subsection{Previous studies of ram pressure and tidal interaction}

There have been numerous studies based on simulations of RP in galaxies \citep[e.g.][]{tonnesen, yun2019, lee2020, akerman23}. Work by \citet{lee2020} showed that the global increase in star formation is higher if the IGM wind is edge-on, rather than face-on. This condition applies to NGC 2276, whose current SFR is 0.74 dex (a factor of $\sim$5) above the main sequence of star formation.

Work by \citet{rasmussen} and \citet{roberts} determined that the velocity of NGC 2276 is supersonic with a value of $\approx$ 900 km/s, causing a bow-shock appearance on the west side.
These two works inferred similar velocities in spite of the very different observational dataset (X-ray vs. radio) and analysis techniques.
They also implied that the main origin of the peculiar morphology in this galaxy comes from ram pressure, supported by the recent findings of \citet{roberts}, where a 100 kpc long tail of synchrotron radiation is reported on the trailing side. The shape and the direction of this tail is consistent with the tail observed in simulations by \citet{lee2020}, and in observations by \citet{Souchereau2025}, which show an expected asymmetry caused by the combined effects of ram pressure and rotation. Moreover, work by \citet{gruendl}, showed that the edge of the gas velocity field on the leading side of NGC 2276, shows an additional redshift of $\sim 20$ km/s, when subtracted from the trailing side, which is consistent with the effect of RP on gas kinematics.

The work of \cite{rnb} claimed that the tail in ram-pressure stripped galaxies does not necessarily point in the direction opposite to the galaxy’s direction of motion, but that the presence of a bow shock, first shown in NGC 2276 by \cite{mulchaey}, would provide a stronger evidence. Considering that both seem to be present in NGC 2276, an east-west motion across the plane of the sky seems to be the most probable.
In contrast to the supersonic velocity of NGC 2276 measured by \citet{rasmussen} and \citet{roberts}, work by \citet{wezgoviec} argued against the existence of a bow shock in NGC 2276 by measuring its velocity to be $\approx 400$ km/s. 
Using XMM-Newton data they concluded that there is no significant change in the temperature of the X-ray emission on the west edge of the disk and the IGM. 
They implied that the shearing forces produced by tidal interactions caused the lopsided morphology of NGC 2276, which highlights how the debate on which effect is predominant remains an open question. However, their claim about the non-existence of the bow shock might be problematic because they did not do analyze the radial distribution of the X-ray emission from the galactic center to the IGM \citep[which was done by][]{rasmussen}. They only measured temperature from the X-ray emission in regions inside the galaxy, outside the leading edge, and far from the galaxy, and thus they do not have its full spatial distribution.

As initially pointed out by \citet{toomre}, tidal interactions are expected to produce a stellar bridge towards the galaxy perturber, and a tail on the opposite side. Additionally, works by \citet{oh08, oh15} claimed that only strong tidal interaction can produce a tail, which begins to appear at about a half orbital time after the bridge started forming. We do not observe a stellar bridge, a tail, nor any strong barring or grand design spirals in NGC 2276 which could indicate that the tidal interaction between NGC 2276 and NGC 2300 is not strong enough to produce it, or that it has not yet passed enough time to form either of them. The former option finds support in \citet{davis96} where they mentioned that NGC 2276 has been within the virial radius of NGC 2300 group for at least 1 Gyr and possibly much longer. Additionally, the tidal disturbances observed around NGC 2300 \citep[e.g.][]{forbes,ilina} could originate from previous mergers, and may not necessarily indicate a recent interaction with NGC 2276.

Findings of elongated asymmetric polarized intensity distribution, and $\sim 2$ times above average total magnetic field intensity in NGC 2276 by \citet{hummel} hinted towards tidal interaction being the main morphological disturber of this system. However, in a later review by \citet{beck2015}, it was claimed that various asymmetric magnetic field distributions could be assigned as a result of both hydrodynamic and gravitational effects.

Furthermore, \citet{wolter15} investigated possible tidal effects in NGC 2276 using a N-body+smoothed-particle hydrodynamical simulation. NGC 2276 was put on a parabolic orbit around NGC 2300, which was assumed to be at the bottom of the gravitational potential well of the group. From this experiment, they concluded that the tidal interaction caused some formation of the tidal arms in both gaseous and stellar components, but that the ram pressure is mostly responsible for the intricate morphology of NGC 2276.

\subsection{Our results - indication of ram pressure as a potential winner?}
\label{sec: disc:rp}

In this work we analysed NGC 2276 with unprecedented high-resolution imaging provided by HST WFC3/UVIS, finding new evidence that favours the scenario of RPS as the dominant mechanism in action.

The first major result is the stellar asymmetry, traced by broadband WFC3/UVIS, and IRAC 1 and 2 filters on the leading and trailing side of the galaxy, visible in both the old stellar disk and the disk of young stars, as shown in Fig. \ref{fig:rad_dist}. 
Bluer filters trace younger stars, and their asymmetry, which closely follows that visible in the ionized gas traced by H$\alpha$, can be attributed to the effects of the edge-on RPS. Considering that, with respect to the old stellar population traced by the NIR bands, the younger stars are more extended on the trailing side, and are more truncated on the leading side (steeper downturn towards red colors at $\sim5.5$ kpc), we believe that these peculiarities regarding the general shape of NGC 2276 can be mainly understood in terms of RPS \citep{boselli}. On the other hand, the asymmetry visible in the reddest filter can potentially be supported with the argument of a previous gravitational interaction \citep[e.g.][]{lokas22} with NGC 2300, although we claim that either the gravitational influence is rather weak, as there are no other visible tidal features, or there were some in the past while NGC 2276 was closer to NGC 2300.

However, work by \citet{smith} revised the idea that RPS affects the gas content alone, leaving the old stars unperturbed. They found that the variation in the gravitational potential induced by the displacement of the ISM can have a visible impact on the distribution and kinematics of the stars and the dark matter. 
This scenario, where the stars are "dragged" by the ram-pressured ISM, might suffice to explain the lopsidedness visible in the redder filters. However, this drag effect would cause the centre of the disk to be dragged downwind by a different amount for different ages of stellar populations, which is not supported by the observations. 

A key caveat to note is that the old disk's morphology in our work was studied using NIR filters from HST and Spitzer. Because young massive stars are significantly brighter than older stars, these filters inevitably include some emission from young stars. On the other hand, the NUV filters are solely sensitive to emission coming from young stars.
A more detailed separation between stellar populations with different ages would require a dedicated spectral energy distribution (SED) modelling analysis that goes beyond the purpose of this work, but which is already planned for our follow-up study.

Additionally, we explored the stellar distribution on the sides of the galaxy that rotate into the IGM (sectors 3 and 4), and with the IGM (sectors 7 and 8) shown in Fig. \ref{fig:rad_dist_48}. 
The results show that the distribution of younger stars seem to be closer to the center in the side that rotates into the IGM (sectors 3 and 4) which could be interpreted as a signature of ram pressure pushing the gas inward, and leaving the old stellar disk relatively unperturbed. 
By observing the color-radial profile of the side that rotates with the IGM (sectors 7 and 8), the distribution of younger stars seems to be farther from the center, which is in agreement with numerical simulations from \citet{rnb} and \citet{jachym09} that showed that the side rotating with the IGM wind has higher degree of RPS.
When comparing the surface brightness profiles, and the overall shape of the galaxy to the simulations \citep[e.g.][]{lee2020, akerman23} there are qualitative similarities that strengthen the evidence of influence of ram pressure in this galaxy. 
The final piece of evidence concerning the distribution of stellar populations that goes in favour of ram pressure is the formation of clumps of young stars, shown in Fig. \ref{fig:RGB}, and visible in Fig. \ref{fig:cont} which could indicate a potential extra-planar star formation induced by ram pressure \citep[e.g.][]{poggianti19, giunchi}, although considering that the orientation of the galaxy is mainly face-on, it is difficult to claim that these clumps are indeed extra-planar, and are rather extended beyond the outer edges of the disk.

Lastly, we investigated the potential spiral arm unwinding in NGC 2276.
Our Fig. \ref{fig:pitch_ang} shows a very clean evidence for unwinding in the spiral arm pattern of NGC 2276. 
Our interpretation that the unwinding is caused by RPS follows the work of \citet{bellhouse}, whose analysis was focused on a sample of cluster jellyfish galaxies where the main ongoing environmental mechanism is RPS. 
By comparing our Fig. \ref{fig:pitch_ang} to work by \citet{bellhouse} (see their Fig. 6) we find similar radial growth of the pitch angle among their various galaxies, again supporting ram pressure as main cause of the lopsided morphology in this galaxy. The higher pitch angle is somewhat more visible on the trailing side, but that interpretation could be biased because we avoided tracing the spiral arms on the front. 

However, it is still possible that tidal interactions play a role in the unwinding mechanism. This option was explored using numerical simulations by \citet{oh08,oh15}, who found that tidal interactions with a low-mass perturber cause an initial growth of the pitch angle with radius (unwinding) during the peak of the interaction, followed by an exponential decay of the pitch angle after $\sim$1 Gyr from the interaction peak. These numerical results cannot be applied directly to the case of NGC 2276, whose supposed perturber (NGC 2300) is not a low-mass system. However, they could indicate that, if the unwinding is gravitationally-driven, the interaction between the two galaxies must have peaked less than 1 Gyr ago.

\section{Summary}
\label{sec: sum}

In this work, we combined newly reduced, publicly available HST WFC3/UVIS images with other ancillary datasets to investigate the role of ram pressure and tidal interactions in shaping the recent evolution of spiral galaxy NGC 2276, a member of the NGC 2300 galaxy group.

In this study we have found two main pieces of evidence that suggest that ram pressure is mainly responsible for the peculiar properties of this galaxy. 
The first is that the surface brightness asymmetry between the leading and the trailing sides of the galaxy is much more pronounced in the bluest HST bands and in the H$\alpha$ image than in the reddest/NIR filters, which we interpret as due to recent ($<$100 Myr) star formation that is displaced from the gravitational centre and compressed on the leading side by ram pressure. While the asymmetry in the reddest bands can be supported with a tidal origin, we must take into account that the broadband filters we used still contain the light from the bright, young stars. We speculate that it is mainly produced by ram pressure, due to compression on the leading side, showing a relatively unperturbed old stellar disk beyond the gas disk, and the bluer extension of the trailing side.
The second piece of evidence is the unwinding of the newly traced 17 overdensity features that build the spiral structure of NGC 2276 (Fig. \ref{fig:pitch_ang}). Following \citet{bellhouse} we interpret this as due to the ram pressure. One possible caveat is that the pitch angle seems to increase primarily on the trailing side, but the detailed tracing of the spiral structure on the leading side was impossible due to the intertwining of the spiral features.

There are two caveats associated with our results. First, it is unclear whether tidal interactions can also contribute to the unwinding. Second, and most importantly, the distinction between the old and the young stellar populations is purely based on broad-band filters, whose light encompasses multiple wavelengths and in turn stellar populations of varying ages, hence it is imperfect.

We plan to address these concerns in our next study via a combination of spatially resolved SED fitting techniques using photometrical and IFU data, which will allow us to robustly disentangle stellar population of different ages, and dedicated hydrodynamical simulations, which are key to quantify the impact of tidal interactions between NGC 2276 and NGC 2300.

\begin{acknowledgements}
We thank the referees for their valuable comments that helped to improve this paper.
This work was supported by the Croatian Science Foundation under the project number [HRZZ-MOBDOK-2023-8006].
This research has made use of the NASA/IPAC Infrared Science Archive, which is funded by the National Aeronautics and Space Administration and operated by the California Institute of Technology.  
This work is also based on observations collected at the Centro Astronómico Hispano-Alemán (CAHA), operated jointly by the Max-Planck Institut für Astronomie and the Instituto de Astrofísica de Andalucía (CSIC). 
A.I. acknowledges the European Research Council (ERC) programme (grant agreement No. 833824, PI: B. Poggianti).
R.S. acknowledges financial support from FONDECYT Regular 2023 project No. 1230441 and also gratefully acknowledges financial support from ANID - MILENIO NCN2024\_112. A.E.L. acknowledges support from the INAF GO grant 2023 ``Identifying ram pressure induced unwinding arms in cluster spirals'' (P.I. B. Vulcani).
This research made use of APLpy \citep{aplpy19}, an open-source plotting package for Python, and Numpy \citep{numpy}.
\end{acknowledgements}

%
%

\bibliographystyle{aa}
\bibliography{bibliography}
\end{document}